\documentclass[aps,prl,twocolumn,groupedaddress,showpacs]{revtex4} 

\usepackage{graphicx}
\usepackage{longtable}

\begin{document}

\renewcommand{\vec}[1]{\mbox{\boldmath$#1$}}

\title{ Prediction of a surface magnetic moment in $\alpha$-uranium}

\author{N.\ Stoji\' c,$^{\ 1,2}$ J.\ W.\  Davenport,$^{\ 2}$ M.\ Komelj,$^{\ 3}$ and  J.\ Glimm$^{\ 2,4}$ }
\affiliation{$^{\ 1}$ Department of Physics and Astronomy,
    Stony Brook University,
        Stony Brook, NY 11794-3381 }

\affiliation{ $^{\ 2}$  Center for Data Intensive Computing, 
    Brookhaven National Laboratory,
    Building 463B,
    Upton, NY 11973-5751}

\affiliation{$^{\ 3}$ Jo\v zef Stefan Institute, 
        Jamova 39,
        SI-1000 Ljubljana, Slovenia }

\affiliation{$^{\ 4}$ Department of Applied Mathematics and Statistics,
        Stony Brook University,
               Stony Brook, NY 11794-3600} 
\date{\today}

\begin{abstract}
Recently, there has been an increased interest in first-principles calculations of the
actinides as well as in finding the new materials which display surface magnetism.
We predict the existence of a magnetic moment on the uranium (001) surface by performing
density functional calculations for a slab geometry in the generalized gradient and 
local spin density approximations with included spin orbit coupling. 
The ferromagnetic phase is energetically favored for all geometries. 
 The calculated total magnetic moment, $0.65{\mu_{B}}$, is 
stable on films of different thickness and it should be observable experimentally.

\end{abstract}

\maketitle

\vspace{5mm}
The beginnings of the field of surface magnetism can be traced back to the end of the 1960s with the claims of 
magnetic dead layers of some transition elements. In subsequent years it has been shown both experimentally and 
by first-principles calculations  that Fe(001), Ni(001) and Cr(001) have considerable enhancement of magnetic moments 
at the surface with respect to the bulk value, 
\cite{OhnFreWei83}.
As a rule one expects the tendency toward magnetism to be increased near metal surfaces, because of the narrowing of 
the density of states which yields a Stoner enhancement in the susceptibility.  This lead to the search for surface 
magnetism in materials which are nonmagnetic in the bulk. There have been many attempts to find such materials 
(vanadium, for example), but most of them proved unsuccessful.
To our knowledge the only experimentally clear case of a magnetic surface on a nonmagnetic metal is 
Rh(100), \cite{WuGarBeg94, GolBarCom99}. 
In contrast, there are many examples of predicted  transition metal monolayers on noble-metal substrates 
such as: Ti, V, Tc, Ru, and Rh  on Ag or Au, \cite{AsaBilHan99}.

 Uranium would appear to be a possible candidate for surface magnetism since there is a rather large change in 
density of states from surface to bulk, 
as pointed out by Hao  ${\it et\ al.}$  \cite{HaoEriFer93}. Also, a comparatively modest expansion of the 
lattice leads to a spontaneous moment 
in the bulk which we have verified, \cite{HjeEriJoh93}.
 In addition, 
 it is near the transition (which occurs beyond plutonium) in which the 5f electrons become
 localized, \cite{Hec01}.  We also note that  many U
 alloys are magnetic, indicating the possible presence of a nearby magnetic instability.  For example, 
URhAl \cite{KunNovDiv01} and UPtAl \cite{AndDivJav01} are ferromagnetic
 with spin moments of order 1$\mu_{B}$. In these structures the U atoms are in hexagonal planes 
(which include the transition element) and are separated by planes with no U.

\begin{figure}[ht] 
\includegraphics[width=6cm]{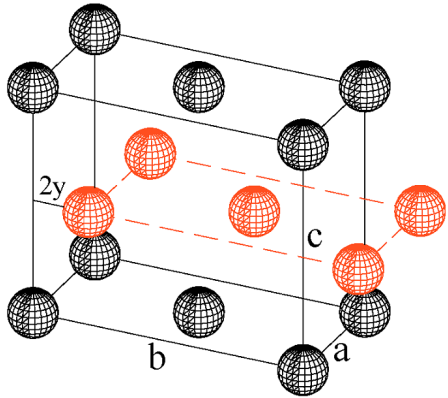}
\caption{$\alpha$ -uranium structure.  }
\bigskip
\label{fig:structure} 
\end{figure}

We have explored the possibility for magnetism in uranium within the framework of the density 
functional theory (DFT). For uranium and the lighter actinides, DFT has been successful at predicting 
structural properties \cite{JonBoeAlb00, Sod02, YamHas90}. For plutonium
 (and presumably for the heavier actinides) calculations which go beyond DFT (such as dynamical 
mean field theory) are required \cite{SavKotAbr01}.  
The calculations have been done using the orthorhombic structure ($\alpha$-uranium), Fig.\ref{fig:structure}. 
This is the stable structure between 43K and 940K.
At higher temperatures, it passes through tetragonal and then bcc structures
 before melting at 1408K.  Below 43K the lattice undergoes a charge density wave (CDW) distortion \cite{MarLanSma90}.
The CDW distortions are only $\approx$0.027 \AA, \cite{MarLanSma90}, and were neglected.
   The orthorhombic structure may be viewed as a distorted
 bcc structure with the (001) plane similar to bcc (110) and with alternate (110) planes shifted along the $b$ axis by 
$(1/2-2y)b$. A similar transformation was described by Axe ${\it et\ al.}$ in \cite{AxeGruLan94}. The bcc structure 
would have $y=$1/4, $a={a_{bcc}}$,
 and $b=c=\sqrt{2}{a_{bcc}}$. At the same volume $a_{bcc}$=3.405 \AA.
 The orthorhombic structure has $a=$2.836 \AA, $b=$5.866 \AA, $c=$4.935 \AA, and $y=$0.1017, \cite{BarMueHit63}. 
It belongs to the non-symmorphic space group Cmcm. We confirmed previous calculations, \cite{Sod02},  which
 show that this structure is lower in energy than bcc by 0.22 eV.  We have also found that the bcc structure is 
mechanically unstable in the sense that the bcc energy is a
 relative maximum with respect to tetragonal distortions (a phonon calculation would lead to negative 
frequencies), \cite{unpublished}. For the calculations reported here we use the
 unrelaxed experimental geometry.

 In this paper we solve the DFT
 equations using the WIEN2k, \cite{BlaSchLui99}, implementation of the full potential linear augmented 
plane wave (FLAPW) method,  \cite{WimKraWei81}, in a supercell geometry. To simulate a surface, we varied 
the length of the vacuum layer until no changes were noticed in energy and density of states and  we kept it at 18 Bohr. 
Our free-standing monolayer calculations were done using 2 atoms per unit cell on
a (001) surface, experimental lattice constants, 150 k points in the irreducible Brillouin zone and
energy cutoff 12.5 Ry. For the film calculations (1 atom per surface cell), number of k-points varied from 306 for the three-layer film 
to 153 for the seven-layer film.
The muffin tin radius was kept fixed at 2.55 Bohr.
For the exchange and correlation we used both the generalized gradient (GGA), \cite{PerBurErn96}, and the local 
spin density  approximations (LSDA) for the monolayer
and only GGA for the film calculations.
 In the actinides it is important to include several corrections to the usual scalar relativistic formulation of 
the FLAPW method.  The spin orbit splitting of the 5f level in the
 isolated atom is 0.82 eV.  
This is comparable, though smaller than the f bandwidth which is of order 2 eV.  Hence it is important to 
include spin orbit effects explicitly.  
This was done treating the spin orbit coupling in a second variation, \cite{BlaSchLui99}. Our bulk energy bands 
and density of states are in
good agreement with fully relativistic results,  \cite{YamHas90}.
In addition, the semicore 6s and 6p levels have bandwidths in the solid of order 2 eV so it is important to include 
them in the basis.  
In the Wien code they are treated by adding local orbitals at the appropriate energies, \cite{MadBlaSch01}.

\begin{table}[ht]
\begin{tabular}{l c c c c } 
\hline
\hline
$ $ & GGA & GGA +SOC & LSDA & LSDA + SOC\\
\hline
FM & $-0.333$ & $-0.226$ & $-0.241$ & $-0.163$ \\
AFM &  $-0.206$ & $-0.136$  & $-0.141$ & $-0.089$ \\
\hline
\hline
\end{tabular}
\caption{Calculated energy differences from the paramagnetic phase for the ferromagnetic (FM) and antiferromagnetic (AFM)
state with and without spin-orbit coupling in GGA and LSDA for a monolayer in eV/atom.}
\label{energies}
\end{table}

\vspace{6mm}
\begin{table}[t]
\begin{tabular}{l c c c c} 
\hline
\hline
$ $ & GGA & GGA+SOC & LSDA & LSDA + SOC\\
\hline
FM - sphere & 2.42 & 2.11 (1.27) & 2.13 & 1.97 (1.13)\\
FM - interstitial & 0.85 & 0.74 & 0.71 & 0.62\\
FM - total per atom  & 3.27 & 2.01 & 2.84 & 1.75\\
AFM - total per atom &  2.25 & 1.92 (0.79)  & 1.89 & 1.69 (0.59) \\
\hline
\hline
\end{tabular}
\caption{Calculated spin (spin $+$ orbital) magnetic moments on spheres and in the interstitial region for the FM and AFM
state with and without spin-orbit coupling in GGA and LSDA for a monolayer in units of $\mu_{B}$. Interstitial contribution
is given per atom. }
\label{magn_moments}
\end{table}

We find that the surface does indeed support a magnetic moment. 
Our results for a monolayer, presented in Table \ref{energies}, demonstrate that the ferromagnetic (FM) phase
is favored over antiferromagnetic (AFM) and paramagnetic (PM) phases. For the GGA with spin-orbit coupling 
(SOC) included, it lies 0.23 eV below the
PM phase and 0.09 eV below the AFM phase. This result  qualitatively does not depend on the exchange-correlation
potential and inclusion of SOC. The direction of magnetization in the case with SOC is [001], 
perpendicular to the surface. The values of spin and total (orbital moment taken into account) magnetic 
moments on spheres and in the interstitial region  
for the FM and AFM phase are
given in  Table \ref{magn_moments}. (Interstitial contributions come out as a consequence of the
LAPW method's partitioning of space into atomic spheres and the interstitial region). 
We can see that the total moments per atom on the uranium monolayer are surprisingly large, 2.01 and 
1.75 $\mu_{B}$ in GGA and LSDA, respectively. Total moment per atom is obtained by adding spin and orbital moments from
a sphere to the spin interstitial moment. The orbital moments in the interstitial region vanish.
We also compare the magnitude of the magnetic moments for the case
without SOC. The moments are significantly larger in that case, similar to
  $4d$ and $5d$ (Ru, Rt, Pd, Os, Ir, Pt$\ldots$) mono and double layers on Ag(001) and Au(001),  \cite{UjfSzuWei95}.

Magnetic moments are similar 
in GGA and LSDA approximations. The influence of the type of the exchange and correlation
approximation on the calculations of magnetic effects on surfaces is a subject of a discussion.
For example, for chromium (001),  \cite{BilAsaBlu00},
it was found that GGA tends to overestimate magnetic moments in 
bulk and at surface. However, the use of the 
LSDA for the description of a magnetic state also gave debatable results (bulk iron, for instance). 
All results presented further in this work
have been obtained using the GGA.

The results for thick films 
are given in Table \ref{layers}.
Since the calculated surface moment is stable upon the change in the slab thickness, we have performed
calculations for up to seven layers.
The  constant value of the magnetic moment  for the three cases is 0.65$\mu_{B}$. 
We show in Table \ref{layers} the spin and orbital moments in the outermost sphere and in the
interstitial region for films of varying thickness.
Interstitial moments are constant for the films of different thickness, from which we
conclude that they are mostly at the surface. They are given per total calculational cell, so dividing them 
by two (two surfaces) and adding to the sum of spin and orbital moment, we get the total
magnetic moment per atom. 
 The moments diminish rapidly as a function of
distance from the surface, so that, for example, for the five-layer film the total  moments
on the spheres are: 0.46, -0.03, and -0.04$\mu_{B}$.
The orbital moments are relatively large. This is 
consistent with the findings for uranium compounds, \cite{KunNovDiv01}.
In Table \ref{energies_film} we present the energy differences from the PM to the FM phase for different
films.

\begin{table}[htb] 
\begin{tabular}{ l c c c c c } 
\hline
\hline
$ $ & 1 layer & 3 layers & 5 layers &  7 layers  \\
\hline
spin moment  & 2.11 & 0.84  & 0.84 & 0.84  \\
orbital moment & $-0.38$ & $-0.38$ & $-0.38$ & $-0.38$ \\
interstitial moment & 0.74 & 0.39 & 0.38 & 0.38 \\
total moment  & 2.01 & 0.66 & 0.65 & 0.65 \\
\hline
\hline
\end{tabular}
 
\caption{Spin and orbital moments on spheres and in the interstitial region at the outermost layer 
for slabs of different thickness, with spin-orbit coupling, given in $\mu_{B}$. For 3 -- 7 layers the total 
moment is taken to be spin $+$ orbital $+$ 1/2 interstitial.}
\label{layers}
\end{table}

\begin{table}[htb] 
\begin{tabular}{ c c c c c } 
\hline
\hline
 1 layer & 3 layers & 5 layers &  7 layers  \\
\hline
  $-0.226$ & $-0.011$ & $-0.016$ & $-0.020$  \\
\hline
\hline
\end{tabular}
\caption{Energy differences from the paramagnetic state for different slab thickness geometries
in eV/atom.}
\label{energies_film}
\end{table}

In order to justify our use of the unrelaxed experimental geometry in this work, 
 we contracted the surface layer of a 5-layer
slab.  The energy minimum occurs at the contraction of 2.66\%. We used the muffin tin radius
of 2.35 Bohr and 72 k points in the irreducible Brillouin zone.
The spin moment on a sphere of the relaxed
structure was reduced slightly to 0.65$\mu_{B}$.

We also calculated the surface energy using the unrelaxed geometry and including the SOC.
It was found to be 0.12 eV/\AA$^{2}$.
This is in a resonable agreement
with the only available experimental result of 0.094eV/\AA$^{2}$ \cite{CRC92}. Previous
calculations of Kollar {\it et al.}, \cite{KolVitSkr94}, gave 0.16 eV/\AA$^{2}$.

\begin{figure}[ht]
\includegraphics[width=7cm]{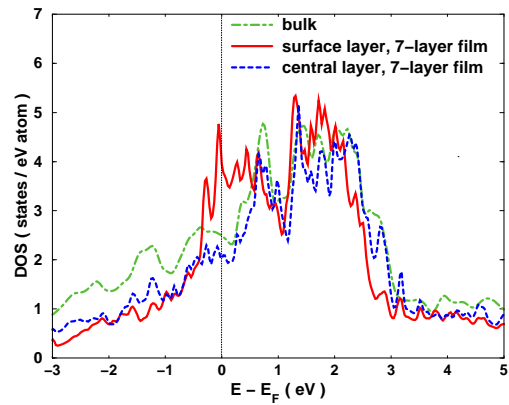}
\caption{Density of states for paramagnetic 7-layer film and bulk.}
\bigskip
\label{fig:DOS_7layer} 
\end{figure}

\begin{figure}[ht]
\includegraphics[width=7cm]{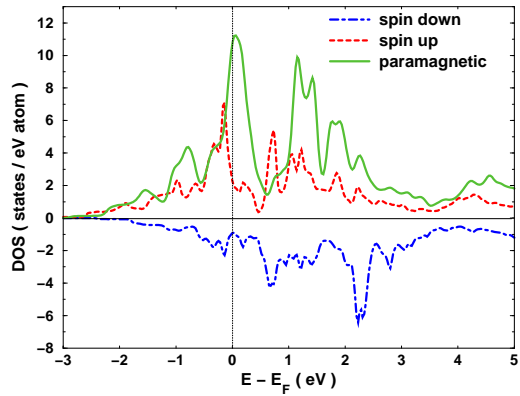}
\caption{Density of states for paramagnetic and magnetic monolayer.}
\bigskip
\label{fig:DOS_monolayer} 
\end{figure}

The existence of the surface magnetic moment can be ascribed to 
the reduction of the coordination number which causes
narrowing and enhancing of the density of states at the Fermi level. This increase in
the local density of states, $n_{loc}(E_{F})$, results in fulfilling the Stoner criterion
$n_{loc}(E_{F}) I > 1$, ($I$ is the exchange integral). For reasonable values of I ($\approx$ 0.2 -- 0.4 eV),
\cite{Jan77}, this change in $n_{loc}$ is sufficient to drive the system magnetic, 
as we observe.
This can be seen from the density of states 
for a paramagnetic 7-layer film and bulk shown in Fig. \ref{fig:DOS_7layer}. The  local density
of states at the Fermi level changes from 2.45  for bulk to 4.01 states/eV  for the surface layer.
It is even more pronounced  for a monolayer, where PM local 
density of states at Fermi level reaches the value of 11.38 states/eV, Fig. \ref{fig:DOS_monolayer}.
A similar effect of an emerging magnetic moment can be produced by a uniform expansion
of the original lattice.  For example, we have
found that there is  a magnetic moment 
of 0.1$\mu_{B}$ at the volume of 26.5$\AA^{3}$ ($\approx $30\% increase).
The moments increase with the further expansion.
 
\begin{figure}[ht] 
\includegraphics[width=7cm]{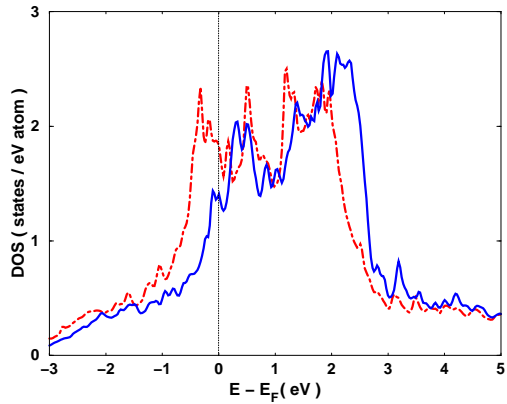}
\caption{Spin-polarized density of states for the outermost layer of the 7-layer film. The solid blue line
represents the spin down density of states and the dot-dashed red line spin up density od states.}
\bigskip
\label{fig:DOS_7sp} 
\end{figure}

In  Fig.\ref{fig:DOS_monolayer}  we show the
spin-polarized density of states for a monolayer, which has a very sharp difference between
spin up and down states. Figure \ref{fig:DOS_7sp}  shows the spin-polarized density of states,
for the surface layer of the 7-layer film.

Experimentally, the surface magnetism can be detected in a variety of ways including, 
for example, spin-polarized photoemission \cite{WuGarBeg94}, linear or circular 
dichroism \cite{GolBarCom99} and surface magneto-optic Kerr
effect \cite{QiuBad99}. Given the size of our computed moments, these effects should be readily 
observable.

N. S. would like to thank E. Vescovo for many interesting and helpful discussions.
 Supported by US Department of Energy under Contract No. DE-AC02-98CH10886.

\def\refname{\ }


\begin{thebibliography} {99}
\bibliographystyle{plain}

\bibitem{OhnFreWei83} See, for example: S.\ Ohnishi, A.\ J.\ Freeman, and M.\ Weinert, {\em Phys. Rev. B}  \/ {\bf 28}, 6741 (1983).

\bibitem{WuGarBeg94} S.\ C.\ Wu, K.\ Garrison, A.\ M.\ Begley, F.\ Jona, and P.\ D.\ Johnson,  {\em Phys. Rev. B} \/ {\bf 49}, 14081 (1994).

\bibitem{GolBarCom99} A.\ Goldoni, A.\ Baraldi, G.\ Comelli, S.\ Lizzit, and G.\ Paolucci, {\em Phys. Rev. Lett.} 
         \/ {\bf 82}, 3156 (1999). 

\bibitem{AsaBilHan99} See, for example: T.\ Asada, G.\ Bihlmayer, S.\ Handschuh, S.\ Heinze, Ph.\ Kurz, and S.\ Bl\" ugel,
         {\em J. Phys. Condens. Matter} \/ {\bf 11} 9347 (1999).

\bibitem{HaoEriFer93} Y.\ G.\ Hao, O.\ Eriksson,  G.\ W.\ Fernando, and B.\ R.\ Cooper, {\em Phys. Rev. B }
        \/ {\bf 47}, 6680 (1993).

\bibitem{HjeEriJoh93} A.\ Hjelm, O.\ Eriksson, and B.\ Johansson, {\em Phys. Rev. Lett.} \/  {\bf 71}, 1459 (1993).

\bibitem{Hec01} S.\ S.\ Hecker, {\em MRS Bulletin} \/ {\bf 26} 672 (2001). 

\bibitem{KunNovDiv01} J.\ Kunes, P.\ Novak, M.\ Divis, and P.\ M.\ Oppeneer, {\em Phys. Rev. B} \/ {\bf 63}, 205111 (2001). 

\bibitem{AndDivJav01} A.\ V. Andreev, M.\ Divis, P.\ Javorsky, K.\ Prokes, V.\ Sechovsky, J. \ Kunes, and Y.\ Shiokawa, 
                      {\em Phys. Rev. B } \/ {\bf 64}, 144408 (2001). 

\bibitem{JonBoeAlb00} M.\ D.\ Jones, J.\ C.\ Boettger, R.\ C.\ Albers, and D.\ J.\ Singh, {\em Phys. Rev. B}
        \/ {\bf 61}, 4644 (2000).  

\bibitem{Sod02} P.\ S\"oderlind, {\em Phys. Rev. B} \/ {\bf 66}, 085113 (2002).

\bibitem{YamHas90} H.\ Yamagami and A.\ Hasegawa, {\em J. Phys. Soc. Jpn. } \/ {\bf 59}, 2426 (1990).

\bibitem{SavKotAbr01} S.\ Y.\ Savrasov, G.\ Kotliar, and E.\ Abrahams, {\em Nature} \/ {\bf 410}, 793 (2001). 

\bibitem{MarLanSma90} J.\ C.\ Marmeggi, G.\ H.\ Lander, S.\ van Smaalen, T.\ Bruckel, and C.\ M.\ E.\ Zeyen, 
                    {\em Phys. Rev. B } \/{\bf 42}, 9365 (1990). 

\bibitem{AxeGruLan94} J.\ D.\ Axe, G.\ Grubel, and G.\ H.\ Lander, {\em J. Alloys and Compounds} \/{\bf 213/214}, 262 (1994).

\bibitem{BarMueHit63} C.\ S.\ Barrett, M.\ H.\ Mueller, and R.\ L.\ Hitterman, {\em Phys. Rev. }\/ {\bf 129}, 625 (1963). 
 
\bibitem{unpublished}  N.\ Stoji\' c, J.\ Davenport, and J.\ Glimm (unpublished).

\bibitem{BlaSchLui99} P.\ Blaha, K.\ Schwarz, G.\ Madsen, D.\ Kvasnicka and J.\ Luitz, {\em WIEN2k, An Augmented Plane Wave + Local Orbitals 
 Program for Calculating Crystal Properties}, (Karlheinz Schwarz, Technical Universit\"at Wien, Austria, 2001).
ISBN 3-9501031-1-2.

\bibitem{WimKraWei81} E.\ Wimmer, H.\ Krakauer, M.\ Weinert, and A.\ J.\ Freeman,  {\em Phys. Rev. B } \/{\bf 24}, 864 (1981). 

\bibitem{PerBurErn96} J.\ P.\ Perdew, K.\ Burke, and M.\ Ernzerhof, {\em Phys. Rev. Let. }\/ {\bf 77}, 3865 (1996).

\bibitem{MadBlaSch01} G.\ K.\ H.\ Madsen, P.\ Blaha, K.\ Schwarz, E.\ Sj\" ostedt, and L.\ Nordstr\" om,
                    {\em Phys. Rev. B } \/{\bf 64}, 195134 (2001). 
 
\bibitem{UjfSzuWei95} B.\ \' Ujfalussy, L.\ Szunyogh, and P.\ Weinberger, {\em Phys. Rev. B}, \/ {\bf 51}, 12836 (1995).
 
\bibitem{BilAsaBlu00} G.\ Bihlmayer, T.\ Asada, and S.\ Bl\" ugel, {\em Phys. Rev. B}, \/ {\bf 62}, R11937 (2000).

 
\bibitem{CRC92}  {\em CRC Handbook of Chemistry and Physics}, (CRC Press, Boca Raton, 1992),
                  $73^{rd}$ edition, p.{\bf 4}-141.

\bibitem{KolVitSkr94} J.\ Kollar, L.\ Vitos, and H.\ L.\ Skriver,   {\em Phys. Rev. B}, \/ {\bf 49}, 11288 (1994).

\bibitem{RosLam68} J.\ W.\ Ross and D.\ J.\ Lam, {\em Phys. Rev.}, \/ {\bf 165}, 311 (1968).

\bibitem{Jan77} J.\ F.\ Janak, {\em Phys. Rev. B}, \/ {\bf 16}, 255, (1977).

\bibitem{QiuBad99} Z.\ Q.\ Qiu and S.\ D.\ Bader, {\em J. Magn. Magn. Mat.}, \/ {\bf 200}, 664 (1999).
 
\end{thebibliography}
\end{document}